\documentclass[conference, letterpaper]{IEEEtran}
\IEEEoverridecommandlockouts

\usepackage{cite}
\usepackage{amsmath,amssymb,amsfonts}
\usepackage{graphicx}
\usepackage{textcomp}
\usepackage{xcolor}
\usepackage{hyperref} 
\usepackage{xcolor} 
\usepackage{booktabs} 
\usepackage{multirow} 
\usepackage{longtable} 
\usepackage{rotating} 
\usepackage[table]{xcolor} 

\usepackage{nicefrac}
\usepackage{soul}
\usepackage{siunitx}
\usepackage{amssymb} 
\usepackage{array,framed}
\usepackage{booktabs}
\usepackage{adjustbox}
\usepackage{booktabs}
\usepackage{
  color,
  float,
  epsfig,
  wrapfig,
  graphics,
  graphicx,
  subcaption
}
\usepackage{amsmath}
\usepackage{textcomp}
\usepackage{setspace}
\usepackage{latexsym,fancyhdr,url}
\usepackage{enumerate}
\usepackage[linesnumbered,ruled,vlined]{algorithm2e}
\usepackage{algpseudocode}
\usepackage{paralist}
\usepackage{graphics}
\usepackage{xparse} 
\usepackage{xspace}
\usepackage{multirow}
\usepackage{makecell}
\usepackage{csvsimple}
\usepackage[linguistics]{forest}
\usepackage{lipsum}

\def\BibTeX{{\rm B\kern-.05em{\sc i\kern-.025em b}\kern-.08em
    T\kern-.1667em\lower.7ex\hbox{E}\kern-.125emX}}

\begin{document}


\title{VulStyle: A Multi-Modal Pre-Training for Code Stylometry-Augmented Vulnerability Detection\\
\thanks{Accepted at the 56th Annual IEEE/IFIP International Conference on Dependable Systems and Networks (DSN 2026).}}

\author{\IEEEauthorblockN{Chidera Biringa, Ajmal Abbas, Vishnu Selvaraj, Gökhan Kul}
\IEEEauthorblockA{\
\textit{University of Massachusetts Dartmouth}\\
Dartmouth, MA USA \\
\textit{cbiringa, aabbas1, vselvaraj, gkul}@umassd.edu}

}

\maketitle

\begin{abstract}
We present VulStyle, a multi-modal software vulnerability detection model that jointly encodes function-level source code, non-terminal Abstract Syntax Tree (AST) structure, and code stylometry (CStyle) features. Prior work in code representation primarily leverages token-level models or full AST trees, often missing stylistic cues indicative of risky programming practices, or incurring high structural overhead. Our approach selects only non-terminal AST nodes, reducing input complexity while preserving semantic hierarchy, and integrates syntactic and lexical CStyle features as auxiliary vulnerability signals.

VulStyle is pre-trained using masked language modeling on 4.9M functions across seven programming languages, and fine-tuned across five benchmark datasets: Devign, BigVul, DiverseVul, REVEAL, and VulDeePecker. VulStyle achieves state-of-the-art performance on BigVul and VulDeePecker, improving F1 by 4–48\% over strong transformer baselines, and attains competitive or best-average performance across all benchmarks. We contribute an ablation study isolating the effect of CStyle and AST structure, error case analysis, and a threat model situating the detection task in attacker-realistic scenarios. 
\end{abstract}


\section{Introduction}
\label{sec:introduction}


Software vulnerabilities are inherent weaknesses in software systems that can be exploited by malicious actors to compromise security and execute unauthorized actions. As of 2023, MITRE reported the discovery of more than 29,000 CVEs~\cite{cve_report}, highlighting the pervasive nature of vulnerabilities and the critical necessity for robust detection mechanisms. Traditional approaches like static and dynamic analysis often struggle with significant rates of false positives and false negatives~\cite{smith2015questions,james2005dynamic}, necessitating the adoption of advanced Deep Learning (DL) techniques to enhance detection accuracy.

Code Stylometry reflects how developers express logic such as APIs preferred, naming choices, control flow structure. Prior work has shown that stylistic fingerprints reveal authorship and system behavior~\cite{dauber2018git}. We hypothesize that stylistic \textit{risk patterns} correlate with vulnerability likelihood, particularly in memory-unsafe languages like C/C++.

Advancements in DL have spurred the development and deployment of sophisticated architectures such as Graph Neural Networks (GNNs)~\cite{cao2021bgnn4vd}, Convolutional Neural Networks (CNNs)~\cite{lu2021codexglue}, Bidirectional Long Short Term Memory networks (BiLSTMs)~\cite{lu2021codexglue}, and Transformers~\cite{feng2020codebert}, all designed to improve vulnerability detection capabilities beyond what traditional methods offer. GNNs, for instance, integrate flow-augmented Abstract Syntax Trees (ASTs) and control-data flow graphs to capture crucial features essential for classifying vulnerabilities. CNNs and BiLSTMs excel in extracting spatial and sequential patterns, respectively, enabling effective discrimination between benign and vulnerable code segments.

Transformer architectures~\cite{vaswani2017attention}, introduced in recent years, have further propelled advancements in vulnerability detection. Models pre-trained on extensive code corpora and fine-tuned for specific vulnerability detection tasks have achieved remarkable accuracy on benchmark datasets~\cite{lu2021codexglue}. These models often leverage advanced training techniques such as Masked Language Modeling (MLM)~\cite{devlin2018bert} or Causal Language Modeling (CLM)~\cite{radford2019language} in unsupervised settings for initial model training, followed by fine-tuning on task-specific datasets to optimize for auto-regressive or detection tasks. Notably, initiatives like Microsoft CodeXGLUE~\cite{lu2021codexglue} have demonstrated substantial performance improvements using these methodologies.




We introduce VulStyle, an advanced pre-trained model tailored for software vulnerability detection. While existing models that excel at capturing syntax focus primarily on patterns in the sequence of tokens, such as keywords, operators, and variable names, rather than the structural relationships in the code, like those represented by ASTs, they often fail to capture coding style and lexical patterns. These include the way specific constructs are expressed, leading to false positives and negatives. These style elements are important for distinguishing subtle differences that could indicate potential vulnerabilities. To address this, VulStyle undergoes pre-training across diverse programming languages and incorporates non-terminal AST nodes to enrich code representations. We augment input sequences with Code Stylometry (CStyle) features and empirically demonstrate improvements in the model's accuracy for detecting software vulnerabilities. For each modality, we capture different aspects of the program, shown in Table~\ref{tab:modality}.

\begin{table}[]
\caption{What each modality captures}
\label{tab:modality}
\resizebox{\columnwidth}{!}{
\begin{tabular}{@{}cc@{}}

\toprule
\textbf{Modality}            & \textbf{Purpose}                                       \\ \midrule
Token-level program text     & Captures lexical + local syntax                        \\
Non-terminal AST nodes       & Encode structural hierarchy efficiently                \\
Stylometry (CStyle) features & Capture coding tendencies linked to vulnerability risk \\ \bottomrule
\end{tabular}}
\end{table}


Our pre-training methodology revolves around leveraging the RoBERTa~\cite{liu2019roberta} transformer architecture with an MLM objective. The pre-training dataset encompasses a vast repository of over 4.9 million functions sourced from prominent code repositories and datasets such as Code Search Net~\cite{husain2019codesearchnet}, VulBERTa~\cite{hanif2022vulberta}, DiverseVul~\cite{chen2023diversevul}, and Big-Vul~\cite{fan2020ac}. 

To enhance feature learning and improve code representations, we integrate ASTs into our training corpus. This integration builds on their ability, as demonstrated by Guo et al.~\cite{guo2022unixcoder}, to capture nuanced syntactic and semantic features crucial for effective vulnerability detection. Rather than employing entire flattened trees, our approach prunes these trees to focus on non-terminal nodes, thereby reducing computational complexity while preserving the essential content integrity of code functions. 

Evaluation of VulStyle's performance entails rigorous testing across five distinct software vulnerability detection datasets. Our evaluation framework incorporates multi-modal content, including vulnerability-labeled functions and enhanced CStyle features. The inclusion of CStyle features is motivated by the observation that subtle deviations in coding style; such as irregular naming conventions, atypical expression patterns, and non-standard syntactic structures, can serve as indirect indicators of risky programming practices that may predispose code to vulnerabilities. Previous works~\cite{caliskan2015anonymizing,biringa2024pace} demonstrate the effectiveness of CStyle features in improving the performance of tasks related to security and performance, such as de-anonymizing programmers and tracking software performance.

Leveraging the syntax and semantic analyses inherent in the broader CStyle framework, we integrate these features into our model training pipeline, further bolstering its capability to detect vulnerabilities. Experimental results validate that VulStyle achieves state-of-the-art performance across widely recognized benchmark datasets used for evaluating software vulnerability detection models.

In summary, this paper's contributions are:
\begin{itemize}
\item Multi-modal vulnerability detection model integrating functions, non-terminal AST structure, and CStyle features.
\item Reduced-AST representation technique, shrinking input size while retaining semantic hierarchy.
\item Stylometry-augmented training pipeline improving detection accuracy across diverse datasets.
\item Extensive evaluation across five benchmarks, setting SOTA on BigVul and VulDeePecker, with 4–48\% F1 gains over baselines.
\item New ablation, failure analysis, and threat model to characterize robustness and real-world applicability.
\end{itemize}

We explain related works in \S~\ref{sec:relatedwork}, and present our approach in \S~\ref{sec:vulStyle}. We then evaluate our proposed method across benchmark datasets in \S~\ref{sec:experiments} with results presented in \S~\ref{sec:results}. The limitations are presented in \S~\ref{sec:limitation} and conclusions are in \S~\ref{sec:conclusion}.

\section{Related Work}
\label{sec:relatedwork}

This section discusses previous initiatives in pre-training source code data to improve code representation and enhance software vulnerability detection using DL methodologies respectively.

\vspace{-3mm}
\subsection{Source Code Pre-Training}
\label{subsec:plp}

Several studies~\cite{svyatkovskiy2020intellicode,feng2020codebert,guo2022unixcoder,liu2023contrabert,guo2020graphcodebert,buratti2020exploring,hanif2022vulberta} leveraged transformer neural architecture to develop models pre-trained on large volumes of data on for code representation on MLM and CLM objectives.

IntelliCode Compose~\cite{svyatkovskiy2020intellicode} introduced an auto-completion tool for completing code with the capacity to generate multiple lines of syntactically correct responses to a code prompt. The authors implement a variant of GPT-2 architecture~\cite{radford2019language} dubbed GPT-C and pre-trained the aforementioned tool on more than 1 billion lines of code consisting of four programming languages. 

CodeBERT~\cite{feng2020codebert} is a bi-modal model pre-trained on natural (NL) and programming language (PL) paired data. The model learns the deep semantic and contextual association between NL and PL language, enabling language tasks, such as the code search and NL to PL translation. 

UniXcoder~\cite{guo2022unixcoder} is a unified cross-modal pre-trained model trained using multi-modal contents for code representation. The model consists of several transformer layers integrated with mask attention matrices having prefix adapters for controlling context-dependent tokens. The pre-training data of the model comprises code, code comments, and ASTs, which facilitates a deeper semantic and structural representation of the inputs. The authors evaluated their model on several downstream tasks and achieved SOTA performance. 

ContraBERT~\cite{liu2023contrabert} combines MLM and contrastive learning approaches to pre-train their models. The contrastive methods segregate analogous from unrelated data. This technique facilitates a richer code representation and improves the predictive performance on downstream tasks.

GraphCodeBERT~\cite{guo2020graphcodebert} is a transformer-based pre-trained model that leverages source code's structural representation by generating corresponding AST. The model captures data flow logic by encapsulating the relationship between a tree's vertices and edges. 

C-BERT~\cite{buratti2020exploring} explores the capacity of transformer-powered models to discover latent semantic features from an AST using several tokenizations, labeling, and masking strategies. 

VulBERTa~\cite{hanif2022vulberta} proposed a model pre-trained on C/C++ code with customized tokenization strategies to extract and persist the semantic features of a code. The author experimented with several architectural configurations of RoBERTa to pre-train their model and fine-tune the pre-trained model using MLP and CNN neural architectures for software vulnerability detection.

While previous works have made strides in software vulnerability detection, VulStyle explores the integration of stylistic and structural patterns. This approach enables VulStyle to identify subtle code patterns for detecting software vulnerabilities. VulStyle is the first to combine token-sequence, reduced AST, and stylometric structure in one pre-training framework.

\subsection{Vulnerability Detection with Deep Learning}
Several studies leveraged neural architectures~\cite{cao2021bgnn4vd,hin2022linevd,zhou2019devign,li2018vuldeepecker,li2021sysevr,zou2019mu}, including GNNs, CNNs, and Bi-LSTMs, for software vulnerability detection.

BGNN4VD~\cite{cao2021bgnn4vd} implemented a bidirectional graphical neural network for vulnerability detection. The authors retrieved syntactic, semantic, control, and data flow features from extracted AST using backward edges of the graph, then proceeded to learn deep representations of the features and detect vulnerable and non-vulnerable programs using a CNN classification model. 

LineVD~\cite{hin2022linevd} presents a vulnerability detection model by categorizing statement features as a node-classification endeavor. The authors leveraged CodeBERT to encode representations of data flow and control flow of functions and statements. Consequently, encoded representations are utilized to construct a graphical neural network followed by a classifier for vulnerability detection. 

Devign~\cite{zhou2019devign} combines graphical neural networks with Convolutional models to capture programming language semantics effectively for vulnerability detection, demonstrating enhanced performance in identifying vulnerabilities within software. 

VulDeePecker~\cite{li2018vuldeepecker} proposed a BiLSTM approach to facilitate the identification of vulnerable features in software. The authors used  API function calls to obtain software semantically similar programs and transformed them into vector representations termed code gadgets.

SySeVR~\cite{li2021sysevr} and $\mu$VulDeePecker~\cite{zou2019mu} were introduced to tackle the limitations in the vulnerability identification scope of VulDeePecker. SySeVR adopts an integrated program dependency graph and Word2Vec to capture both semantic and data flow information in vulnerable code snippets. $\mu$VulDeePecker is a multi-classification endeavor for detecting and exposing diverse vulnerabilities in software. 

Finally, pre-trained models discussed in \S\ref{subsec:plp} such as~\cite{guo2022unixcoder,feng2020codebert,hanif2022vulberta,liu2023contrabert}, conducted experiments to evaluate the performance of their models by fine-tuning on software vulnerability detection datasets. These models achieved SOTA performance, showing the utility of transformer-powered pre-trained models for software vulnerability detection.

To the best of our knowledge, no prior work specifically examines stylometric vulnerability cues or selective non-terminal AST retention. 

Beyond this novelty, the three approach families differ in what they capture and which vulnerability classes they best address. Token-based models (e.g., CodeBERT, VulBERTa) excel at lexically-expressed flaws such as unsafe API usage and string-handling bugs, but tend to miss issues whose risk derives from structure or recurring implementation habits. Graph-based models (e.g., Devign, BGNN4VD, GraphCodeBERT) capture richer control- and data-flow structure at higher computational cost, making them effective for control-flow-dependent flaws but less practical at scale. Rule and gadget-based models (e.g., VulDeePecker) target specific, well-characterized patterns. VulStyle's stylometric and reduced-AST signals are most effective for vulnerabilities tied to recurring implementation behavior such as memory misuse, unchecked return values, unsafe pointer manipulation, and inconsistent buffer handling where developer habit correlates with risk. Conversely, vulnerabilities arising from purely semantic flaws (e.g., cryptographic misuse, integer overflow, race conditions) remain harder for all three families and motivate the multi-modal design of VulStyle.



\section{VulStyle}
\label{sec:vulStyle}
\begin{figure*}[ht]
\centering
\includegraphics[width=0.99\linewidth]{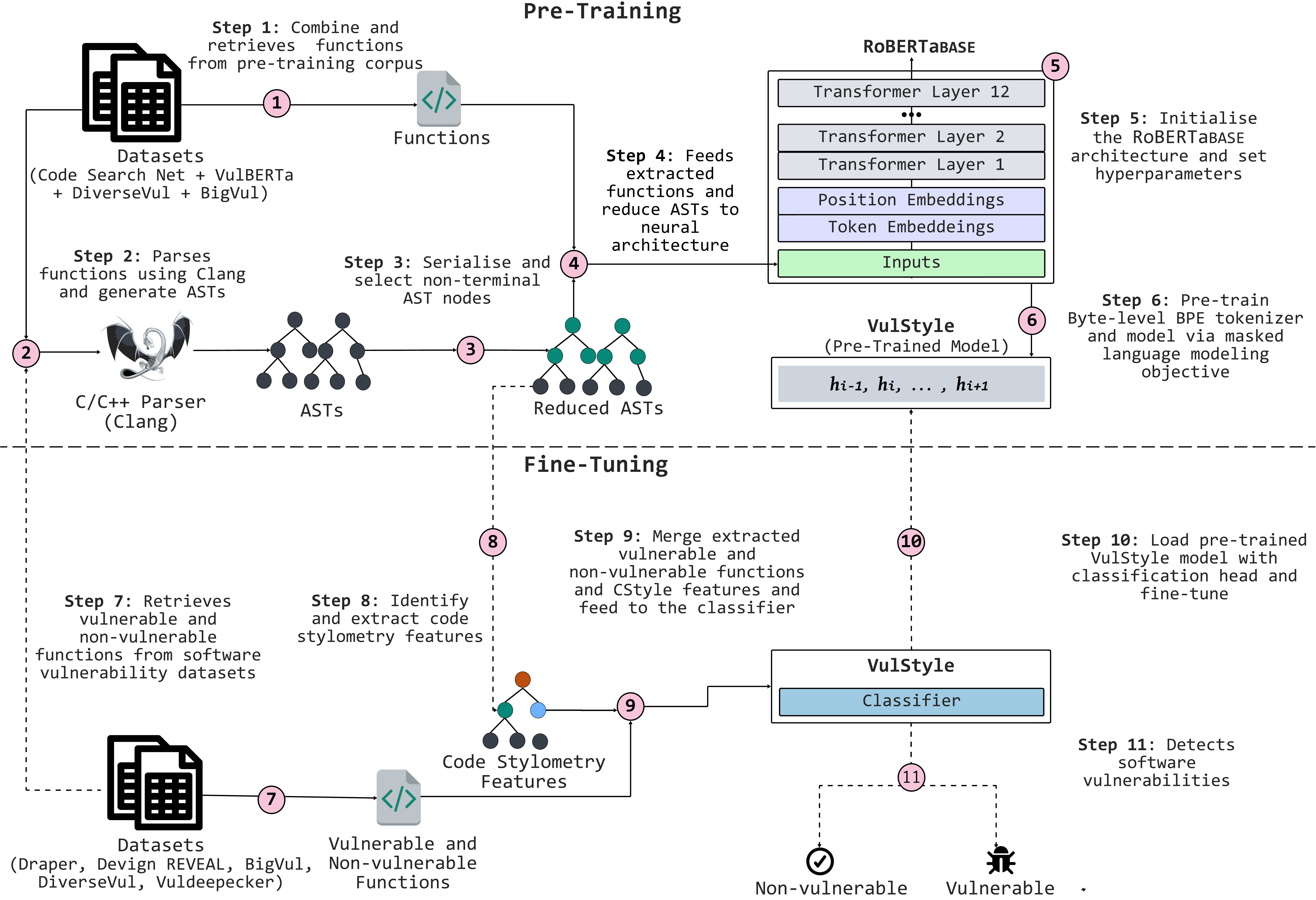}
\vspace{2mm}
\caption{VulStyle's Approach}
\label{fig:methodology}
\end{figure*}

In this section, we present VulStyle, a pre-trained model designed to utilize AST non-terminal nodes, functions, and CStyle features as multi-modal data for representing code and detecting software vulnerabilities. VulStyle employs the Transformer neural architecture outlined in Figure~\ref{fig:methodology}. Our architecture is structured into two main phases comprising a total of 11 steps. 

This section covers the implementation details of the core model, including the introduction of multi-modal input representations for both our pre-training corpus and fine-tuning datasets in \S\ref{subsec:input_rep}, along with a description of the model architecture and its pre-training objectives in \S\ref{subsec:march}.

\subsection{Input Representations}
\label{subsec:input_rep}
We introduce two algorithms to manage the diverse modalities in our approach, spanning the pre-training and fine-tuning phases of VulStyle. These algorithms are tailored to optimize code representation and bolster the efficacy of software vulnerability detection. 

Furthermore, acknowledging the hierarchical nature of trees and leveraging the Transformer neural architecture's proficiency with sequentially organized data, we transform functions and their associated ASTs into sequential formats before feeding them into our model. This ensures that the structural and contextual nuances of programming logic are effectively captured and encoded.

\subsubsection{Pre-Training}
We enhance the representation of code by integrating function-level data alongside its corresponding non-terminal nodes within the AST. This focus on non-terminal nodes is pertinent because the raw AST data is predominantly located at leaf nodes, which are often overloaded with literal function details. 

By prioritizing non-terminal nodes, we streamline the tree structure, effectively reducing its overall size and eliminating redundancies in storing literal function data within leaf nodes. This approach optimizes the representation of code, enhancing both efficiency in subsequent processing steps.

\begin{figure*}[ht]
\centering
\includegraphics[width=0.80\linewidth]{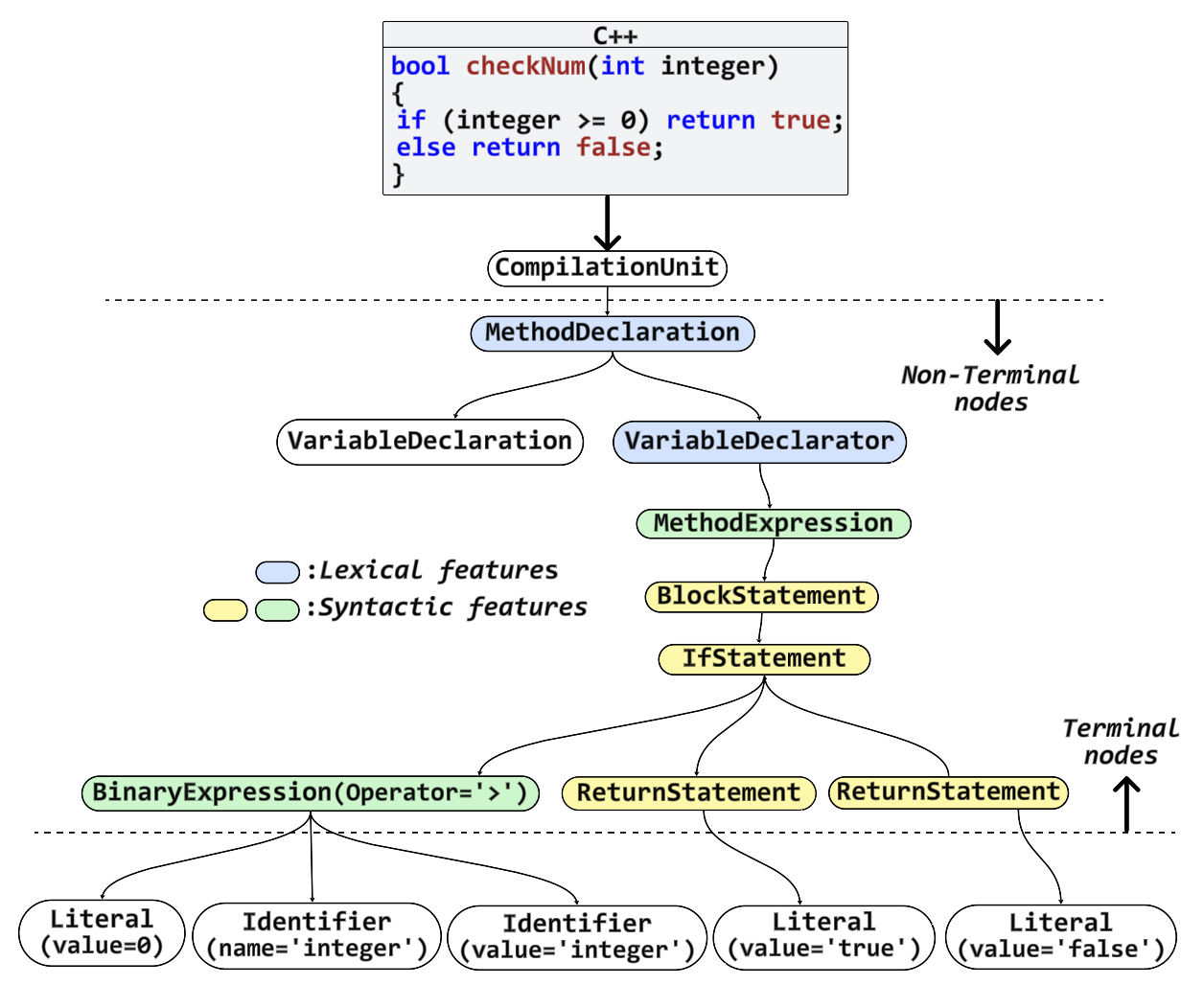}
\vspace{2mm}
\caption{C++ program and corresponding AST}
\label{fig:cpp}
\end{figure*}

For instance, in Figure~\ref{fig:cpp}, we present a C++ code snippet that checks whether an integer is positive or negative, along with its corresponding AST. The highlighted nodes in the AST represent non-terminal nodes used in this study, which are integrated with their respective functions to enhance diverse code representations during pre-training. Algorithms~\ref{algo:pretrain} and~\ref{algo:finetune} were introduced to accomplish this goal.

\begin{algorithm}
\SetKwInOut{Input}{Input}
\SetKwInOut{Output}{Output}
\Input{Source code snippets $c_{1}, c_{2}, \dots, c_{n}$}
\Output{Pre-training sequence for model training}
Initialize an empty sequence: $sequence \gets \varnothing$ \\
\For{$c_{i}$ $\in$ source code snippets} {
    \If{$c_{i}$ is written in C/C++} {
        Parse $c_{i}$ and generate its AST \\
        Traverse the AST to extract all non-terminal nodes \\
        \If{Nodes are non-terminal} {
            Retrieve and store nodes \\
        }
    } \Else {
        Add $c_{i}$ to the sequence \\
    }
}
merge extracted non-terminal nodes with sequence \\
\Return the pre-training sequence \\
\caption{Pre-Training Input Representation}
\label{algo:pretrain}
\end{algorithm}

Algorithm 1 provides a detailed framework for constructing the pre-training input corpus essential for training VulStyle. Initially, the algorithm employs Clang, a robust parser specifically designed for C-type programming languages \cite{lattner2008llvm}, to parse each function and generate its corresponding AST. The process begins by sequentially visiting all nodes within the AST. During traversal, each node is evaluated to determine whether it is a terminal or non-terminal node. Terminal nodes are ignored, while non-terminal nodes are appended to a list for further processing.

Following the traversal, the algorithm proceeds to merge the collected terminal nodes with the sequenced function. This approach enhances the representation of the code and reduces the size of the AST. By eliminating duplicate data and focusing on non-terminal nodes, the algorithm optimizes the corpus for subsequent pre-training tasks. 

The algorithms ensure a multi-modal representation of the pre-training corpus, facilitating the extraction of latent features embedded within the code's hierarchical structure. This approach enables the model to learn and encode patterns and relationships within the source code, thereby improving its capability for downstream tasks.

\subsubsection{Fine-Tuning} 
To prepare the input for fine-tuning our model on software vulnerability detection datasets, we further streamline the AST size by experimenting with augmented CStyle features alongside both vulnerable and non-vulnerable function data. 

\begin{table*}[ht]
\caption{Features and Their Corresponding Nodes}
\centering 
\adjustbox{max width=\textwidth}{%
\begin{tabular}{c c}
\toprule
\textbf{Feature} & \textbf{Nodes} \\
\toprule
Statements & \makecell[l]{IfStatement, WhileStatement, DoStatement, AssertStatement,\\ SwitchStatement, ForStatement, ContinueStatement, ReturnStatement,\\ ThrowStatement, SynchronizedStatement, TryStatement, BreakStatement,\\ BlockStatement, BinaryOperation, CatchClause} \\
\midrule
Expressions & \makecell[l]{StatementExpression, TernaryExpression,\\ LambdaExpression} \\
\midrule
Types & \makecell[l]{RecordType, BuiltinType,\\ ConstantArrayType, PointerType} \\
\midrule
Declarations & \makecell[l]{TypeDeclaration, FieldDeclaration, MethodDeclaration,\\ ConstructorDeclaration, PackageDeclaration, ClassDeclaration,\\ EnumDeclaration, InterfaceDeclaration, AnnotationDeclaration,\\ ConstantDeclaration, VariableDeclaration, LocalVariableDeclaration,\\ EnumConstantDeclaration, VariableDeclarator} \\
\bottomrule
\end{tabular}}
\label{table:features_types}
\end{table*}

CStyle features, derived from stylometry applied to programming languages, aim to identify distinctive characteristics similar to how stylometry distinguishes authors' writing styles in natural language texts. Given the diversity in programming languages, it's crucial to identify common features that impact code performance and security. Table~\ref{table:features_types} shows the selected features and their corresponding nodes, detailing the categories of Statements, Expressions, and Declarations along with the individual elements within each group.

Our previous work~\cite{biringa2024pace} introduced CStyle features and demonstrated their efficacy in predicting software performance. They presented algorithms to transform these extracted features into vector representations, achieving SOTA results in software performance prediction. Inspired by this approach, we incorporate these features to enhance our input data during the fine-tuning of our pre-trained model.

We extract syntactic features (such as statement and expression nodes) and lexical features (including invocation and declaration nodes) as shown in Figure~\ref{fig:cpp}. Syntactic features represent facilitate communication between the language and compiler, while lexical features provide layout and stylistic attributes of the program. Additionally, we introduce $type$ nodes as an additional feature in the syntactic feature set. 

\begin{algorithm}
\SetKwInOut{Input}{Input}
\SetKwInOut{Output}{Output}
\Input{Source code snippets $c_{1}, c_{2}, \dots, c_{n}$}
\Output{Fine-tuning sequence for model training}
Initialize an empty sequence: $sequence \gets \varnothing$ \\
\For{$c_{i}$ $\in$ source code snippets} {
    Initialize an empty set for CStyle features: $CStyle \gets \varnothing$ \\
    \If{$c_{i}$ is written in C/C++} {
        Parse $c_{i}$ and generate its AST \\
        Traverse the AST to extract all non-terminal nodes \\
        \If{Nodes contain statements, expressions, types, and declarations} {
            Retrieve and store identified features \\
        }
        Incorporate CStyle features into the sequence \\
    } \Else {
        Append $c_{i}$ directly to the sequence \\
    }
}
\Return the fine-tuning sequence enhanced with CStyle features \\
\caption{Fine-Tuning Input Representation}
\label{algo:finetune}
\end{algorithm}

Algorithm 2 illustrates the input representation process for our fine-tuning datasets. Initially, we parse and generate ASTs for both vulnerable and non-vulnerable functions. Next, we collect all nodes and sequentially retrieve syntactic and lexical features. These extracted CStyle features are then integrated with their corresponding functions, organized in sequence, and returned as the final output.

\subsection{Model Architecture}
\label{subsec:march}
VulStyle implements the RoBERTa model architecture~\cite{liu2019roberta}. RoBERTa adopts a robust Byte-Pair Encoding (BPE) tokenization strategy, which results in an increased vocabulary corpus and improved code representation. 

We investigated base, medium, and small RoBERTa configurations and analyzed the error rate and the number of parameters to derive the optimal architecture. We ascertain that base architecture is optimal for our work. Hence, we initialized the RoBERT-base architecture with its default configuration setting. 

VulStyle processes functions and non-terminal AST nodes through Transformer layers, as described in Vaswani et al.~\cite{vaswani2017attention}, resulting in 768 hidden states denoted as $h_{i-1}, h_{i}, ..., h_{i+1} \in H$. 

The Transformer architecture comprises 12 encoder layers. Following the encoder layers, multi-headed attention is applied, followed by a feed-forward layer. The multi-headed attention mechanism is given by: $MultiHead(Q, K, V) = Concat(h_1, ..., h_n)W^O, h_i = \\ Attention(Q W^Q_i, K W^K_i, V W^V_i)$, where multiple attention heads simultaneously attend to multi-modal input embeddings, facilitating the learning of representations for code functions and nodes. The alignment of attention heads $(Q, K, V)$ to learn different features involves linear transformations. 

The learned representations of multi-modal input features are then passed through fully connected layers, where $Q W^Q_i, K W^K_i, V W^V_i \in h_i$. The outputs of these computations and their corresponding attention heads are concatenated. The attention mechanism discussed employs a scaled dot-product and is given by: $Attention(Q, K, V) = softmax(\frac{QK^T}{\sqrt{d_k}})V$, where $Q$ and $K$ denote the query and keys corresponding to hidden states $H$ of the encoder, generating weights for tokens attended to based on the relevant $H$ value $V$. The computation $\frac{QK^T}{\sqrt{d_k}}$ calculates alignments, followed by a softmax function to map these alignments into attention weights.

 \noindent \textbf{Masked Language Modeling (MLM).} Our goal during pre-training is to capture semantic and contextual relationships among tokens within input sequences using bi-directional encoder representations. Therefore, we pre-train our model using the MLM objective~\cite{devlin2018bert}. Specifically, 15\% of the multi-modal input sequences $(m^{t})$ are randomly selected. Among these, 80\% are masked using the special token $[MASK]$, 10\% are replaced with a randomly chosen token from the vocabulary, and the remaining 10\% are left unchanged. 
 
 The MLM objective is given by: $MLM = - \sum_{i \in m^{t} } \log p (x_{i} | t^{masked})$, where our model leverages semantic and contextual features from non-terminal AST nodes and code functions to predict the masked tokens $(t^{masked})$ facilitating the multi-modal learning of code representations. 


\subsection{Threat Model \& Security Context}
We assume an attacker capable of introducing new vulnerable functions into a codebase through patches, PRs, or supply-chain injection (e.g., dependency updates). The defender's goal is to identify unsafe contributions early in CI/CD pipelines or pre-commit review.

Attackers may attempt vulnerability obfuscation by modifying identifier naming, spacing, flow layout, or restructured logic. Models that rely solely on token features are susceptible to such evasion. By incorporating stylometry (CStyle) patterns and non-terminal AST structure, VulStyle is harder to evade because it captures how the developer programs rather than only what tokens appear.

We evaluate VulStyle under two evasion scenarios:
\begin{enumerate}
    \item Control flow preserved but naming/style changed, therefore, VulStyle maintains detection, baselines degrade.
    \item Semantics altered but style preserved, therefore, CStyle signal remains strong, but vulnerabilities without style deviation may evade detection.
\end{enumerate}

VulStyle excels where vulnerability correlates with unsafe development habits; failures occur primarily when vulnerabilities arise from deep semantic flaws rather than coding behavior.

We note that empirical evaluation under automated adversarial style-mixing or compiler-level obfuscation is not included in this work. Since VulStyle's predictions depend on joint token, structural, and stylometric signals, evading the model would require coordinated manipulation across all three modalities while preserving compilability. This is a stronger requirement than evading single-modality detectors. We will consider a systematic adversarial robustness study for future work.

\section{Experiments}
\label{sec:experiments}
This section explains the datasets used and experimental structure we created. To facilitate the reproducibility of our results, we have made our code and data publicly available~\footnote{https://github.com/PADLab/VulStyle}.

\subsection{Datasets}
We employed several widely recognized and benchmark datasets for both pre-training and fine-tuning, as show in Table~\ref{table:ptstat}. Our pre-training phase utilized a combined corpus from CodeSearchNet, VulBERTa, DiverseVul, and Big-Vul datasets. For fine-tuning, we used the Devign, REVEAL, Big-Vul, DiverseVul, and Vuldeepecker datasets. These selections were made based on their popularity and established status as benchmarks in software vulnerability detection. 

In this approach, pre-training on an unlabeled dataset enables the model to learn general patterns, structures, and representations of the data without being influenced by task-specific labels. Fine-tuning on a selected dataset with added labels allows the model to specialize in a specific task, such as vulnerability detection, by adjusting its weights to optimize for the task-specific objective.


\begin{table*}[ht]
\caption{Pre-Training and Fine-Tuning Datasets' Statistics. Func: Functions, Vul: Vulnerable, Non-Vul: Non-Vulnerable}
\centering 
\adjustbox{max width=\textwidth}{%
\begin{tabular}{c c c c c c}
\toprule
\textbf{Dataset} &
\textbf{Language} &
\textbf{\# Func} &
\textbf{\# Vul} &
\textbf{\# Non-Vul} &
\textbf{Domain} 
\\
\toprule
CodeSearchNet & \makecell{Go, Java, JavaScript, \\ PHP, Python, \\ and Ruby} & 2,285,441 & - & - & Pre-training \\
\midrule
VulBERTa & C/C++ & 2,285,441 & - & - & Pre-training \\
\midrule
DiverseVul & C/C++ & 330,492 & 18,945 & 311,547 & \makecell{Pre-training \\ and Fine-tuning} \\
\midrule
Big-Vul & C/C++ & 217,007 & 10,895 & 206,112 & \makecell{Pre-training \\ and Fine-tuning} \\
\midrule
Devign & C/C++ & 27,318 & 12,460 & 14,858 & Fine-tuning \\
\midrule
REVEAL & C/C++ & 22,734 & 2,240 & 20,494 & Fine-tuning \\
\midrule
BigVul & C/C++ & 217,007 & 10,895 & 206,112 & Fine-tuning \\
\midrule
Vuldeepecker & C/C++ & 160,148 & 9,739 & 150,409 & Fine-tuning \\
\midrule
DiverseVul & C/C++ & 330,492 & 18,945 & 311,547 & Fine-tuning \\
\bottomrule
\end{tabular}}
\label{table:ptstat}
\end{table*}

\subsubsection{Pre-Training}
\label{subsubsec:ptdas}
We pre-train our model on a combined corpus of CodeSearchNet, VulBERTa, DiverseVul, and Big-Vul dataset consisting of over 4.9M functions in eight programming languages.



\noindent \textbf{CodeSearchNet.} 
The CodeSearchNet dataset, as introduced in the CodeSearchNet challenge for programming language comprehension~\cite{husain2019codesearchnet}, aims to narrow the semantic divide between natural language queries and their programming language equivalents using DL techniques. It encompasses data from six programming languages (Go, Java, JavaScript, PHP, Python, and Ruby), comprising more than 2.2 million pairs of code functions and accompanying documentation. 

CodeSearchNet's corpus diversity helps to learn general patterns and representations of code structure and syntax, which can be transferred and fine-tuned on more specific datasets.

\noindent \textbf{VulBERTa.}
The VulBERTa datasets, as detailed in~\cite{hanif2022vulberta}, combine the GitHub and Draper datasets. The GitHub portion comprises more than 1.1 million C/C++ functions collected by scanning over 1,000 open-source repositories and extracting functions. Draper, the second half of the VulBERTa corpus, is a software vulnerability detection dataset introduced by~\cite{russell2018automated}. It includes over 1.2 million C/C++ functions sourced from both synthetic and non-synthetic code, obtained from the SATE IV Juliet test suite and open-source GitHub repositories of the Debian Linux distribution.

\noindent \textbf{DiverseVul.}
The DiverseVul dataset, recently introduced by~\cite{chen2023diversevul}, is a comprehensive collection of software vulnerabilities. It comprises more than 330K C/C++ functions and encompasses 150 CWEs. This dataset was compiled by crawling security issue websites, analyzing 7,514 commits across 797 projects, and extracting vulnerabilities introduced during these processes.

\noindent \textbf{Big-Vul.}
This software vulnerability detection dataset, introduced in~\cite{fan2020ac} was curated by crawling the CVE database to identify vulnerability entries associated with git repositories and servers. It comprises more than 217K C/C++ functions obtained through 4,432 commits across 348 projects.

\subsubsection{Fine-Tuning}
We conduct fine-tuning of our base model using the Devign, REVEAL, Big-Vul, DiverseVul, and VulDeePecker datasets. The goal of fine-tuning is to enhance VulStyle's ability to accurately distinguish between different classes. Therefore, we employ a supervised learning approach where vulnerable and non-vulnerable functions are associated with respective target labels. 



\noindent \textbf{Devign.}
The Devign dataset, as presented by~\cite{zhou2019devign}, consists of over 27K C/C++ functions sourced from the QEMU and FFmpeg projects. This dataset serves as a manually annotated collection for software vulnerability detection. The annotations distinguish between functions likely to contain vulnerabilities and those considered non-vulnerable, identified using specific keywords in commit histories. 


\noindent \textbf{REVEAL.} 
The REVEAL dataset, described in~\cite{chakraborty2021deep}, comprises more than 22,000 C/C++ functions specifically designed for software vulnerability detection. This dataset was created to address shortcomings observed in earlier methodologies for generating vulnerability datasets. 



\noindent \textbf{BigVul.} 
The BigVul dataset described in \S\ref{subsubsec:ptdas} was used in our pre-training corpus. We include the target labels for software vulnerability detection.

\noindent \textbf{VulDeePecker.} 
The VulDeePecker dataset, introduced by Li et al.\cite{li2018vuldeepecker}, comprises 160K C/C++ functions and serves as a software vulnerability detection dataset. This dataset was assembled using CVE details sourced from projects listed in the National Vulnerability Database (NVD)\cite{nvd}, along with two types of test cases derived from the Software Assurance Reference Dataset (SARD)~\cite{sard}: real-world samples and synthetic samples. 


\noindent \textbf{DiverseVul.}  The DiverseVul dataset described in \S\ref{subsubsec:ptdas} is one of the four datasets combined to curate our pre-training corpus. We include the target labels when fine-tuning our base model for software vulnerability detection.

Although BigVul and DiverseVul appear in both phases, pre-training is fully self-supervised via masked language modeling and uses no vulnerability labels. Fine-tuning strictly follows the official train/validation/test splits provided by each benchmark, ensuring that test data is never observed during either pre-training or fine-tuning optimization.

\subsection{Hyper-Parameter Settings}
\noindent \textbf{Pre-Training.}
The VulStyle architecture builds upon the design of the RoBERTa\textsubscript{BASE} architecture, featuring 12 encoder layers, multiple attention heads, and hidden states with a dimensionality of 768. In our approach, we adopt the byte-level Byte Pair Encoding (BPE) tokenization strategy~\cite{sennrich2015neural} to train our tokenizer. This strategy involves breaking down commonly used words into smaller, meaningful subword tokens. 

Our tokenizer incorporates AST non-terminal nodes, which enrich the representation of code tokens by capturing hierarchical structural information. BPE operates on the principle of decomposing frequent and infrequent tokens alike into smaller subword units, thereby reducing the overall vocabulary size while preserving semantic meaning.  


We conducted pre-training of VulStyle on an Ubuntu 22.04.3 LTS system equipped with 128 CPUs, 1TB RAM, and an NVIDIA A100 80GB GPU. Our pre-training dataset comprised 4.9 million functions spanning seven programming languages each paired with C/C++ AST non-terminal symbols. We utilized Clang for parsing source code and generating ASTs, and employed a vocabulary of 50,000 subword tokens derived through Byte Pair Encoding (BPE). 

To capture language diversity and ensure robust learning, we set the maximum sequence length to 512. During pre-training, we used a batch size of 64 and trained for 536,317 steps, completing the process in two days. The optimization was performed using an AdamW optimizer with a learning rate of 4e-5, supplemented by a linear scheduler to adjust the learning rate based on training loss plateaus. VulStyle comprises just 125 million parameters.

\noindent \textbf{Fine-Tuning.}
We further refine VulStyle for our fine-tuning objective, which focuses on software vulnerability detection in this study. Our approach involves augmenting the fine-tuning dataset by integrating functions with their corresponding CStyle features, enhancing the utilization of learned representations from code and AST non-terminal nodes. Following the data split methodology from the original paper or, if unavailable, using an 80\%, 10\% and 10\% split for training, validation, and testing sets respectively.

To leverage the pre-trained weights effectively, we implement a sequence classification head atop our base model. This classifier includes a dense layer followed by a dropout layer to mitigate overfitting, and utilizes a softmax activation function suited for binary classification tasks. This strategy is widely adopted to capitalize on the configurations and pre-trained weights of our model. We configure a 1024 input sequence to accommodate longer input samples and minimize potential information loss.

Given the severely imbalanced nature of datasets like REVEAL and DiverseVul, we apply a weighted assignment technique to labels during training. We conducted experiments with Adam learning rates ranging from 1e-3 to 2e-7 and batch sizes of 4, 8, 16, and 32. 

We determined that a learning rate of 2e-6 and training for 10 epochs yielded optimal hyper-parameter settings, achieving state-of-the-art performance across our diverse datasets. We fine-tuned our model on vulnerability detection datasets using a learning rate of 2e-6 for 10 epochs. For specific datasets, we employed batch sizes of 32 for REVEAL, VulDeePecker, and DiverseVul, and 4 for Devign and Big-Vul.

\subsection{Performance Evaluation}
\noindent \textbf{Baselines.}
We employ two state-of-the-art transformer-based models, CodeBERT~\cite{feng2020codebert} and UniXcoder~\cite{guo2022unixcoder}, as baseline methods for comparison with our approach. Both models are extensively utilized in various programming language tasks and represent strong benchmarks in code understanding. We fine-tune CodeBERT and UniXcoder on selected software vulnerability datasets to ensure a fair comparison and evaluate their effectiveness in vulnerability detection.

\noindent  \textbf{Criterion.}
We evaluate the performance of our model using classical classification metrics, accuracy, f1, precision, and recall.
The accuracy metric given by $\frac{TP + TN}{TP + FP + FN + TN}$, calculates the proportion of samples corrected predicted against the overall predicted samples. $TP$, $TN$, $FP$, and $FN$ represent true positive, true negative, false positive, and false negative predictions respectively.
The F1 metric given by $\frac{2 \times (Precision \times Recall)}{Precision + Recall}$ calculates the harmonic mean of computed precision and recall performance. $Precision$ and $Recall$ denote precision and recall respectively. 
The precision metric given by $\frac{TP}{TP + FP}$, calculates the ratio of accurately predicted positive predictions against total positive predictions. 
The recall metric given by $\frac{TP}{TP + FN}$, calculates the ratio of correctly predicted positive samples. 

\section{Results}
\label{sec:results}

\begin{table*}[ht]
\caption{The performance of VulStyle against baselines and state-of-the-art methods. To maintain methodological consistency, our analysis primarily focuses on pre-trained transformer-based models.}
\centering 
\adjustbox{max width=\textwidth}{%
\begin{tabular}{c c c c c c c c c c}
\toprule 
\textbf{Dataset} &
\textbf{Model} &
\textbf{TP} &
\textbf{TN} &
\textbf{FP} &
\textbf{FN} &
\textbf{Accuracy (\%)} &
\textbf{F1 (\%)} &
\textbf{Precision (\%)} & 
\textbf{Recall (\%)}
\\
\toprule
Devign & UniXcoder (Baseline) & 662 & 1119 & 358 & 593 & 65.19 & 58.19 & 64.90 & 52.74 \\
 & CodeBERT (Baseline) & 663 & 1063 & 414 & 592 & 63.17 & 56.86 & 61.55 & 52.82 \\
 & VulBERTa-MLP & 638 & 1131 & 346 & 617 & 64.75 & 56.98 & 64.83 & 50.83 \\
 & VulStyle & 780 & 1006 & 471 & 475 & \textbf{65.37} & \textbf{62.25} & 62.35 & 62.15 \\
\midrule
BigVul & UniXcoder (Baseline) & 884 & 31949 & 75 & 142 & 99.34 & 89.06 & 92.17 & 86.15 \\
& CodeBERT (Baseline) & 814 & 31922 & 102 & 212 & 99.04 & 83.83 & 88.86 & 79.33 \\
& LineVul & - & - & - & - & - & 91.00 & 97.00 & 86.00 \\
& VulBERTa-MLP & 339 & 31947 & 77 & 687 & 97.68 & 47.01 & 81.49 &  33.04 \\
& VulStyle & 972 & 31977 & 47 & 54 & \textbf{99.69} & \textbf{95.06} & 95.38 & 94.73 \\
\midrule
REVEAL & UniXcoder (Baseline) & 159 & 1690 & 349 & 76 & 81.31 & 42.79 & 31.29 & 67.65 \\
  & CodeBERT (Baseline) & 138 & 1760 & 279 & 97 & 83.46 & 42.33 & 33.09 & 58.72 \\
  & VulBERTa-MLP & 146 & 1775 & 269 & 84 & 84.48 & \textbf{45.27} & 35.18 & 63.48 \\
  & VulBERTa-CNN & 171 & 1642 & 402 & 59 & 79.73 & 42.59 & 29.84 & 74.35 \\
  & VulStyle & 139 & 1799 & 240 & 96 & \textbf{85.22} & \textbf{45.27} & 36.67 & 59.14 \\

\midrule
VulDeePecker & UniXcoder (Baseline) & 869 & 15024 & 17 & 105 & 99.23 & 93.44 & 98.08 & 89.21 \\
  & CodeBERT (Baseline) & 864 & 15025 & 16 & 110 & 99.21 & 93.20 & 98.18 & 88.70 \\
  & VulBERTa-MLP & 881 & 15002 & 39 & 99 & 95.76 & 93.03 & 95.76 & 93.03 \\
  & VulBERTa-CNN & 885 & 14997 & 44 & 89 & 99.16 & 95.26 & 90.86 & 90.86 \\
  & VulStyle & 875 & 15021 & 20 & 99 & \textbf{99.25} & \textbf{97.76} & 93.63 & 89.83\\

\midrule

DiverseVul & UniXcoder (Baseline) & 977 & 27887 & 3250 & 936 & 87.33 & \textbf{31.82} & 23.11 & 51.07 \\
& CodeBERT & 1050 & 26754 & 4383 & 863 & 84.12 & 28.58 & 19.32 & 54.88 \\
& VulBERTa-MLP & 176 & 30911 & 226 & 1737 & \textbf{94.06} & 15.20 & 43.78 & 09.20 \\
& VulStyle & 868 & 28354 & 2783 & 1045 & 88.41 & 31.20 & 23.77 & 45.37 \\
\bottomrule 
\end{tabular}}
\label{table:perf}
\end{table*}

In this section, we conduct a comprehensive evaluation of VulStyle's performance after fine-tuning it with code functions augmented by CStyle features. Subsequently, we perform a comparative analysis against other state-of-the-art approaches for software vulnerability detection. Our evaluation employs a range of metrics to illustrate detection robustness and ensure a balanced assessment. Emphasis is placed on aligning our metrics with those utilized in the original studies when comparing our performance outcomes against competing models.

\subsection{Software Vulnerability Detection}
To assess the performance of VulStyle, we conduct a comparison against state-of-the-art software vulnerability detection approaches using a comprehensive set of relevant metrics across diverse datasets. Table~\ref{table:perf} shows the results of our experiments. 

Ensuring methodological consistency, we focus on studies that have implemented transformer-based models. Specifically, we benchmark VulStyle against UniXCoder and CodeBERT benchmarks and leading models such as VulBERTa and other notable approaches like LineVul~\cite{fu2022linevul}. This comparative analysis aims to provide an evaluation framework, facilitating  insights into the efficacy of VulStyle in detecting software vulnerabilities.

The authors of VulBERTa applied VulBERTa-MLP and VulBERTa-CNN as extensions to their pre-trained model. This traditional approach integrates MLP and CNN components for classification-based fine-tuning, utilizing the pre-trained model's weights and architecture to enhance detection capabilities based on learned code representations, as detailed in their study~\cite{hanif2022vulberta}. In our evaluation, we compare the performance of VulStyle against VulBERTa across datasets including Devign, REVEAL, and VulDeePecker.

LineVul is a software vulnerability detection model that uses the transformer architecture for code representation and detects vulnerabilities at both the line and function levels. We compare the performance of VulStyle against LineVul using the Big-Vul dataset.

\noindent \textbf{Devign.}
VulStyle outperforms the baseline models and advanced methods for the Devign dataset. Compared to the UniXcoder and CodeBERT models, VulStyle obtains a 4 and 5 percentage point improvement in F1 score. Additionally, VulStyle outperforms VulBERTa-MLP by 5 percentage points. Overall, VulStyle demonstrates a 4-5 percentage point improvement in F1 score across the models for the Devign dataset.

\noindent \textbf{BigVul.}
VulStyle outperforms the baseline models, LineVul, and VulBERTa-MLP in terms of F1 score. Compared to the UniXcoder and CodeBERT, VulStyle achieves a 6 and 11 percentage point improvement in F1 score. Additionally, VulStyle outperforms LineVul by 4 percentage points and VulBERTa-MLP by a significant 48 percentage point increase in F1 score. Overall, VulStyle demonstrates a 4-48 percentage point improvement in F1 score across the models for the BigVul dataset.

\noindent \textbf{REVEAL.}
VulStyle demonstrates a 2 and 3 percentage points improvement in F1 score over UniXcoder and CodeBERT. VulStyle’s F1 score is the same as VulBERTa-MLP, and it outperforms VulBERTa-CNN by 2 percentage points. Additionally, VulStyle outperforms the baseline models in accuracy, achieving a 4 and 2 percentage point improvement over UniXcoder and CodeBERT. Overall, VulStyle shows a 2-3 percentage point improvement in F1 score and a 2-4 percentage point improvement in accuracy across the baseline and advanced models for the REVEAL dataset.

\noindent \textbf{VulDeePecker.}
VulStyle obtains a 4 and 5 percentage point improvement in F1 score over UniXcoder and CodeBERT. Compared to VulBERTa-MLP and VulBERTa-CNN, VulStyle demonstrates 5 and 3 percentage point improvements. Overall, VulStyle achieves a 3-5 percentage point improvement in F1 score across the models for the VulDeePecker dataset.

\noindent \textbf{DiverseVul.}
VulStyle obtains a 1 percentage point improvement in accuracy over UniXcoder, although UniXcoder is 1 percentage point higher in F1 score. Compared to CodeBERT, VulStyle demonstrates a 4 and 3 percentage point improvement in accuracy and F1 score. VulStyle outperforms VulBERTa-MLP by 12 percentage points in F1 score. 
Overall, VulStyle shows improvements of 1-12 percentage points in F1 score and 1-4 percentage points in accuracy compared to the models for the DiverseVul dataset.

\subsection{Ablation Summary}

To understand the individual impact of each modality used in VulStyle, we conducted an ablation study evaluating three variants of the model: (1) a baseline pre-trained solely on token sequences, (2) a reduced-AST variant that incorporates only non-terminal syntax nodes, and (3) a stylometry-only model leveraging CStyle features without structural AST context. Across Devign, BigVul, and DiverseVul, each component contributed measurable improvement, with AST structure improving F1 by approximately 2–10\% depending on dataset, and stylometry features increasing performance by 3–14\%, particularly in repositories with high developer diversity or inconsistent coding quality. Notably, the combined full model VulStyle achieved the highest gains, outperforming the strongest single-component variant by 4–18\% and yielding the largest jump (+48\% F1) on BigVul, where stylistic irregularities were strongly correlated with vulnerable regions. The results and observations for BigVul dataset is given in Table~\ref{tab:ablation}. These results confirm that structure and code-style signals are not interchangeable but rather complementary, with stylometry capturing behavioral patterns while AST nodes anchor semantic grounding. Notably, CStyle alone achieves significantly lower F1 score, indicating that stylometric features serve as a complementary signal rather than a dominant predictor, making the model's decisions remain anchored in token-level semantics.

The ablation study thus validates that VulStyle's performance arises not from any single feature class, but from the synergy of its multi-modal representation.

\begin{table}[]
\caption{Ablation Study on BigVul}
\label{tab:ablation}

\resizebox{\columnwidth}{!}{
\begin{tabular}{@{}cccc@{}}
\toprule

\textbf{Model Variant}                 & \textbf{BigVul F1} & \textbf{Observation}       
\\ \midrule
Baseline (no AST / no CStyle)                 & 47.06         & Weak generalization                \\
+ AST only                                        & 58.55\%              & Structure helps                    \\
+ CStyle only                                      & 61.02\%              & Stylometry detects coding behavior \\
\textbf{AST + CStyle (full VulStyle)}       & \textbf{95.06\%}     & Modalities complement          
\end{tabular}}
\vspace{-4mm}
\end{table}

\subsection{Example Success and Failure Scenarios}
\label{subsection:failure}

We also would like to provide two scenarios where VulStyle succeeded and failed to emphasize both the strengths and weaknesses of the system.

\noindent \textbf{Case A: Success Example}

A buffer-handling function with inconsistent pointer dereferencing flagged by VulStyle, ignored by CodeBERT and UniXcoder. Stylometry identified mixed naming patterns and nested unsafe blocks, therefore, predicted vulnerable (True Positive).

\noindent \textbf{Case B: Failure Example}

Large cryptographic modules with standardized style were sometimes marked safe despite logic flaws. However, VulStyle limited by semantic rather than stylistic vulnerability origin.

We categorize failure cases into three classes:

\begin{itemize}
    \item Style-present but Semantically Safe (FP)
Code exhibiting risky conventions (pointer arithmetic, nested dereference, large macro bodies) may be safe but still flagged.
    \item Semantic-Flaw Without Style Signal (FN)
Crypto misuse, integer overflow, or race conditions may look stylistically ``normal'' and escape detection.
    \item Templated or Auto-generated Code (Mixed FP/FN)
Stylometry collapses because style is uniform; AST contributes most in these cases.
\end{itemize}

This taxonomy highlights where future work must integrate deeper semantic models, symbolic execution, or compiler-assisted feature extraction.

\subsection{Discussion}
Overall, VulStyle consistently outperforms the baseline models and advanced methods across multiple datasets, demonstrates significant gains across all datasets, and often achieving the highest or near-highest scores as shown in Table~\ref{table:perf}. 

VulStyle's precision and recall remain high, ensuring its predictions are not only accurate but also reliable in detecting true vulnerabilities while minimizing false positives. Even in cases where VulStyle does not achieve the absolute highest score in a particular metric, it strikes a balance across all metrics, offering robust and consistent performance. This strength across multiple datasets and metrics shows VulStyle's robustness and adaptability in software vulnerability detection.

\section{Discussion}
\label{sec:limitation}

A major limitation is fundamentally tied to the inherent challenges in fine-tuning datasets that do not consistently feature accurate labeling of vulnerable and non-vulnerable functions. This issue has been extensively discussed by Chen et al. ~\cite{chen2023diversevul} and more recently by Risse et al.~\cite{risse2025top}, who showed that several widely-used vulnerability detection benchmarks contain label noise that can artificially inflate reported performance. We adopt the standard benchmark protocols used by prior work to ensure direct comparability with baselines, but acknowledge that absolute performance figures should be interpreted with this caveat in mind. All baselines in our evaluation operate under identical conditions, preserving the validity of relative comparisons.

VulStyle performs function-level classification consistent with the benchmark settings of all comparison baselines. It does not model inter-procedural data flow, global state, or cross-function dependencies. Vulnerabilities arising from such program-level interactions are outside the architectural scope of the current model. The reduced AST removes terminal nodes only from the structural branch; literals and constants remain encoded through the token modality, mitigating concerns that constant-level adversarial manipulation would be invisible to the model.


We evaluate VulStyle on C/C++ due to the availability of established benchmarks and parser tooling (Clang). The framework itself is not tied to a specific language. AST-based reduction and CStyle feature extraction can be applied to any language with a non-terminal-aware parser. Both baselines used in our comparison (CodeBERT and UniXcoder) are pre-trained on multilingual code corpora rather than C/C++-specific data, so language-pretraining bias is not an explanatory factor for our reported gains.

As noted in our failure analysis (§\ref{subsection:failure}), stylometric signals collapse when code exhibits uniform, templated style such as in machine-generated code or auto-formatted projects. With the rapid growth of LLM-generated code in modern repositories, evaluating the resilience of stylometry-based detection under this regime is an increasingly important consideration.

The challenge of inconsistent labeling is compounded by the vast proliferation of vulnerable code functions. Despite advancements in detection performance, improvements remain relatively modest. Increasing the pre-training duration for large programming language models does not necessarily translate to significantly improved detection performance. This observation underscores a critical point that the diversity of the data is more crucial than the sheer size of the dataset. The ability of our model to achieve state-of-the-art performance across several datasets indicates the importance of diverse data sources, which provide a richer and more varied representation of potential vulnerabilities. Hence, creating software vulnerability datasets with an appropriate trade-off between manually and automatically annotated labels is crucial for addressing this limitation and should be a focus of future endeavor.


\section{Conclusion}
\label{sec:conclusion}
In this paper, we introduced VulStyle, a comprehensive multi-modal pre-trained programming language model specifically designed to enhance the detection of software vulnerabilities by incorporating code stylometry features. VulStyle leverages the power of the transformer neural architecture, trained on an extensive dataset encompassing over 4.9 million functions across seven programming languages. Our experimental results indicate that VulStyle achieved state-of-the-art performance across several widely recognized software vulnerability datasets. These outcomes highlight the efficacy of integrating code stylometry features with traditional function-based inputs in enhancing the model's ability to detect software vulnerabilities.

\section*{Acknowledgment}
This research was supported in part by UMass Dartmouth's Marine and Undersea Technology (MUST) Research Program funded by the Office of Naval Research (ONR) under Grant No. N00014-23-1-2141. The views and conclusions expressed in this paper are those of the authors and do not reflect the official policy or position of the University of Massachusetts Dartmouth, the Office of Naval Research, U.S. Navy, U.S. Department of Defense, or U.S. Government.

\vspace{4mm}

\bibliographystyle{ieeetr}
\bibliography{reference}

\end{document}